\documentclass[12pt]{article}
\pdfoutput=1
\usepackage{color,amsmath,amssymb,exscale,psfrag,epsfig}
\usepackage{cite,color,url}
\usepackage{pdfpages}
\usepackage{graphicx}
\usepackage{slashed}
\usepackage[utf8]{inputenc}
\usepackage[T1]{fontenc}
\usepackage{xcolor}
\usepackage{accents}
\usepackage{centernot}
\usepackage{enumerate}
\usepackage{subfig}
\usepackage{float}
\usepackage{placeins}
\usepackage{floatpag}

\renewcommand{\H}{{\cal H}}
\usepackage[colorlinks=true
,urlcolor=blue
,anchorcolor=blue
,citecolor=blue
,filecolor=blue
,linkcolor=blue
,menucolor=blue
,linktocpage=true
,pdfproducer=medialab
]{hyperref}
\input epsf
\usepackage{lineno}
\def\bea{\begin{eqnarray}}
\def\eea{\end{eqnarray}}

\definecolor{nicered}{rgb}{0.7,0.1,0.1}
\definecolor{nicegreen}{rgb}{0.1,0.5,0.1}
\hypersetup{colorlinks,citecolor= nicegreen,linkcolor= nicered}

\def\be{\begin{equation}}
\def\te{\end{equation}}
\def\ee{\end{equation}}
\def\ba{\begin{eqnarray}}
\def\bea{\begin{eqnarray}}
\def\nn{\nonumber}
\def\tea{\end{eqnarray}}
\def\ea{\end{eqnarray}}
\def\eea{\end{eqnarray}}

\def\bfra{\begin{frame}}
\def\efra{\end{frame}}
\def\al#1\fal{\begin{align}#1\end{align}}
\def\bfra#1\efra{\begin{frame}#1\end{frame}} 

\def\dd{\tilde{\alpha}}
\def\ddd{\beta}

\catcode`\@=11
\def\lsim{\mathrel{\mathpalette\@versim<}}
\def\gsim{\mathrel{\mathpalette\@versim>}}
\def\@versim#1#2{\vcenter{\offinterlineskip
\ialign{$\m@th#1\hfil##\hfil$\crcr#2\crcr\sim\crcr } }}
\catcode`\@=12
\catcode`@=12
\begin{document}
\thispagestyle{empty}
\begin{flushright}
ICAS 048/20
\end{flushright}
\vspace{0.1in}
\begin{center}
	{\Large \bf A Machine Learning alternative to placebo-controlled clinical trials upon new diseases: A primer } \\
\vspace{0.2in}
{\bf Ezequiel Alvarez$^{(a)\dagger}$,
Federico Lamagna$^{(a,b)\ddag}$,
Manuel Szewc$^{(a)\diamond}$
}
\vspace{0.2in} \\
{\sl $^{(a)}$ International Center for Advanced Studies (ICAS), UNSAM and CONICET,\\
	Campus Miguelete, 25 de Mayo y Francia, (1650) Buenos Aires, Argentina }
\\[1ex]
{\sl $^{(b)}$ Centro At\'omico Bariloche, Instituto Balseiro and CONICET\\
Av.\ Bustillo 9500, 8400, S.\ C.\ de Bariloche, Argentina}
\end{center}
\vspace{0.1in}

\begin{abstract}
	The appearance of a new dangerous and contagious disease requires the development of a drug therapy faster than what is foreseen by usual mechanisms.   Many drug therapy developments consist in investigating through different clinical trials the effects of different specific drug combinations by delivering it into a test group of ill patients, meanwhile a placebo treatment is delivered to the remaining ill patients, known as the control group.  We compare the above technique to a new technique in which all patients receive a different and reasonable combination of drugs and use this outcome to feed a Neural Network.  By averaging out fluctuations and recognizing different patient features, the Neural Network learns the pattern that connects the patients initial state to the outcome of the treatments and therefore can predict the best drug therapy better than the above method.  In contrast to many available works, we do not study any detail of drugs composition nor interaction, but instead pose and solve the problem from a phenomenological point of view, which allows us to compare both methods.  Although the conclusion is reached through mathematical modeling and is stable upon any reasonable model, this is a proof-of-concept that should be studied within other expertises before confronting a real scenario.   All calculations, tools and scripts have been made open source for the community to test, modify or expand it.   Finally it should be mentioned that, although the results presented here are in the context of a new disease in medical sciences, these are useful for any field that requires a experimental technique with a control group.
\end{abstract}

\vspace*{2mm}
\noindent {\footnotesize E-mail:
{\tt 
$\dagger$ \href{mailto:sequi@unsam.edu.ar}{sequi@unsam.edu.ar},
$\ddag$ \href{mailto:federico.lamagna@cab.cnea.gov.ar}{federico.lamagna@cab.cnea.gov.ar},
$\diamond$ \href{mailto:mszewc@unsam.edu.ar}{mszewc@unsam.edu.ar}
}}

%%%%%%%%%%%%%%%%%%%%%%%%%%%%%%%%%%%%%%%%%%%%%%%%%%%
\newpage
\section{Introduction}
\label{section:1}

Current and last decades research in drug discovery, and drug therapy design in clinical trials have implemented Machine Learning (ML) techniques in order to optimize drug design, therapy and disposition.   Drug discovery and design is a key field for tackling new diseases, and in recent years has used molecular structures and target activities databases in combination with ML techniques to learn and discover potential biological active molecules and their corresponding human drug targets \cite{dd1,dd2,dd3,dd4,dd5,dd6}.  Clinical trials on known and new drugs, as well as on combination therapies, is one of the most expensive costs in biopharma and medicine and has also taken profit of ML techniques to enhance its efficacy and reduce its costs.  These techniques have been applied for treatments in a diversity of diseases and recent reviews and relevant papers can be found in Refs.~\cite{ct1,ct2,ct3,ct4,kaptein} and references therein.

Clinical trials are strictly regulated by their corresponding agencies.  However, in many cases of seriously ill patients a compassionate drug therapy is allowed.  Upon the appearance of a new and contagious disease such as the COVID-19 \cite{cv1,cv2} or any other, urgent working drug therapies are required and usual development times must be stretched and optimized.  In many cases a clinical trial consists in a placebo-controlled study: a test group receives a drug treatment and a control group receives placebo.  This is necessary to recognize the real drug effect while comparing it to a placebo treatment.  These trials need many patients in each of the groups in order to reduce the statistic fluctuations as well as fluctuations that may be originated in uncontrolled or unknown variables which may exist in the patients or in each specific treatment.  If the number of patients is large enough, these fluctuations wash-out and the real effect of the specific drug treatment may be analyzed.  The problem with this technique is that each specific drug treatment needs to be taken by a large number of patients, and therefore there is little room to expand in different combinatory treatments.  This is translated into a time dilation in finding the best option in drug combinations that would yield the best available therapy.

Upon the advent of Artificial Intelligence and ML techniques these premises may change.  Although a normal human research requires to test the same drug and a placebo in many patients to learn its effect as the fluctuations average to zero, a complex Neural Network (NN) could in principle learn from a more diverse dataset in which all patients are tested in a different drug combination.  The NN would learn the pattern of the drugs' effect whereas the fluctuations would wash-out to zero in the loss function.  This {\it `reasoning'} of the Neural Network could enlarge the scope of the number of possible drug combinations to be analyzed in comparison to the above described usual clinical trial when both techniques use the same number of patients.  If this is the case, then the ML technique would be an important tool to find a better drug treatment for a new disease.  This is the problem we investigate and present in the current article.

We propose to explore whether with the assistance of Machine Learning techniques there could be better ways than the placebo-controlled studies to tackle the problem of finding the best drug therapy of a new unknown disease.  This work is purely mathematical consisting in modeling and simulations, and therefore does not have other ambition than being a proof-of-concept of the program described below.  Reference \cite{kaptein} considers a similar approach, although we use different tools.

This work is divided as follows.  In Section \ref{problem} we pose the problem and understand it within a specific mathematical language.  In Section \ref{pheno} we propose phenomenological model to address the problem and allow us to perform a diversity of calculations and simulations to show our results.  Section \ref{discussion} contains a few remarks on the model and our results, as well as future prospects that open up from these results.  We conclude in Section \ref{conclusions} and collect some relevant formulas and verification results in Appendices \ref{functions} and \ref{NN}.  All calculations and programs to reproduce and expand the results in this paper can be found as supplementary material in the {\tt GIT} repo in \cite{git}.

\section{Problem Setup}
\label{problem}

Upon the appearance of a new disease, there is a set of possible drugs which can be combined to treat it.  However, the correct or best combination is usually unknown.  The drugs' effect on the patients is, among others,  non-linear --drugs may interfere with one another--, and patient-dependent.   This scenario can be mathematically described as follows.  We consider the patient features and initial conditions with a multidimensional vector $\vec x$, where each component $x_i$ corresponds to any kind of patient features or pre-existing conditions which may be relevant for the study, as for instance diabetes or heart pre-conditions, genomics, age group, ethnicity, etc..  The drugs whose combinations are to be studied are described with a multidimensional vector $\vec y$, where each component corresponds to the daily dose of drug $i$ to be delivered in treatment.  For the sake of concreteness and simplicity, for given patient $\vec x$ and drug combination $\vec y$, we assume a unique treatment that consists in delivering drug combination $\vec y$ every day during a specific fixed number of days.  Therefore, for any $(\vec x, \vec y)$ all treatments consist in the same technique, but with different patient and drug combination.  As a result of this treatment, we assume that the patient outcome can be described with a number between 0 and 1: 0 being dead and 1 being in excellent health conditions.  In an exact science there would be a function that connects the initial condition and features of the patient to the outcome through a given specific treatment.  This health function, which we define as $h(\vec x, \vec y)$, is a multi-variable non-linear unknown function that can take values in $[0,1]$. However, in real scenarios there are many uncontrolled variables which also affect the outcome of the given treatment, as it can be some unknown pre-existing condition, a genomic factor, or different physicians providing different evaluation of the patients, among many others.  These uncontrolled variables could be included as a stochastic noise on $h$ in different ways.  One practical way is to consider a factorizable stochastic modulation on $h$,
\begin{equation}
	\H(\vec x, \vec y) = S(\vec x, \vec y) \,  h (\vec x, \vec y),
	\label{f}
\end{equation}
while either re-normalizing, dropping or overflowing\footnote{ In this work we have used renormalization and overflowing and $S$ independent of $\vec x$ and $\vec y$, as described in more detail in Appendix \ref{functions}.} values outside $[0,1]$.   This noisy function $\H$ is the one with physical meaning that best represents in practice the health outcome of a patient with features $\vec x$ when receiving a treatment consisting in drug therapy $\vec y$.   In this framework we have that 
\begin{equation}
	\H(\vec x,  \vec 0) = \mbox{Health outcome of a patient receiving placebo or no drugs during treatment}
	\nonumber
\end{equation}

Within this mathematical modeling, the problem of finding the best drug treatment for the disease on patient $\vec x$ is translated into finding the drug combination $\vec y$ that maximizes $\H(\vec x, \vec y)$.  In most of this work we are interested in finding the drug therapy that best heals in average a set of patients, therefore in this case we are interested in finding the drug combination $\vec y$ that maximizes the average of $\H(\vec x, \vec y)$ for a set of patients $\vec x$.  Actually, in the cases in which the dependence of $\H$ on the patient features is subleading compared to the dependence on the drug treatment, both maximums would be similar.

Therefore, given $N$ patients we want to explore which of the following techniques would be the best option to find the drug combination $\vec y$ that maximizes the distribution of $\H(\vec x, \vec y)$ over a new given set of patients:
\begin{itemize}
	\item {\bf A) Regular Drug Therapy technique (RDT):} Divide $N$ in $k$ sets of $n$ patients ($n\ \cdot k = N$) and in each one of these sets apply a given drug therapy $\vec y_i$ ($i=1...k$) to half of the patients ($n/2$) and placebo to the other half, also known as the control group.  After the treatment, consider in each set the average outcome of the patients receiving treatment and the average of patients receiving placebo.  (There are many different clinical trials analyses \cite{clinical0, clinical}, however here we adopt one very similar to current COVID-19 clinical trial on Remdesivir \cite{remdesivir}.)  The set that yields the largest difference between these two averages, and its corresponding drug therapy $\vec y_A$, is considered to be the best drug therapy within this technique.  In the previously defined mathematical language, $\vec y_A$ provides the best distribution of $\H(\vec x , \vec y)$ over $\vec x$.  Where {\it best} means that provides the largest average of $\H$, and therefore would provide the best average healing for patients.
	\item {\bf B) Neural Network Drug Therapy on the RDT data (NN@RDT):} Using the above $N$ data points we train a Neural Network (NN) to learn $\H(\vec x, \vec y)$.  Once this NN is trained we simulate a large set of pseudo-patients and drug therapies $(\vec x, \vec y)$, and find the drug therapy $\vec y$ that yields the maximum average for the trained NN output on $(\vec x, \vec y)$.  We define this drug therapy as $\vec y_B$, and estimate that provides the best distribution of $\H(\vec x , \vec y)$ over $\vec x$.
	\item {\bf C) Neural Network Drug Therapy technique (NNDT):} Apply a different drug therapy to each one of the $N$ patients, observe the result after the treatment, and train a NN with this data to make it learn $\H$.  Then, follow the same procedure as in NN@RDT, simulating a large set of pseudo-patients and drug therapies, finding the best drug therapy $\vec y_C$.
\end{itemize}

There are a few points to be discussed about these techniques.  What we name as {\it Regular} is just because this is the technique against which we want to contrast the other two, and because it is currently used in an important clinical trial for COVID-19 \cite{remdesivir}.  There are many other techniques --or improvements of this-- which are currently being used, including some of them using ML techniques.  The technique NN@RDT does not need in principle other data points that those in RDT, although it is very likely that different sets in RDT have a spread in their setups (duration, forms of dosage delivery, etc.).  Technique NNDT can resemble in principle purely theoretical, since in reality one cannot try any drug combination in patients.  However, one can use mathematics to reduce the space of drug therapies $\vec y$ in such a way that only feasible therapies appear as $\vec y$ is varied (see below).  

Finally, it is worth stressing at this point that this work aims to compare the above techniques and determine whether it would be possible to issue an statement that could work for a general class of functions $\H$, and therefore in particular to the true function.  It is not the objective of this work to inquire about the true form of $\H$\footnote{See however discussion in Section \ref{discussion}}.  As a matter of fact, the whole work remains phenomenological and mathematical, with no connection to real drug compounds nor real patient features.

\section{Phenomenological approach}
\label{pheno}

Along this section, we address the posed problem from the phenomenological point of view.  We do not consider the intermediate steps, nor inquire on the real chemical or biological reactions which have been and are extensively studied, instead we consider only the initial condition and the final outcome of the patient\footnote{A similar approach with different tools is found in Ref.~\cite{kaptein}, we discuss our main differences in Section \ref{discussion}}.  Also, we do not consider the individual meaning of the variables $(\vec x, \vec y)$, but we only require $\vec x$ to be features and initial conditions of the patient, and $\vec y$ any part of the treatment that can be varied at will.  As a matter of fact the components of $\vec y$ could have other non-drug meaning that physicians consider relevant for the treatment and the whole procedure would still be valid. 

\begin{figure}[h!]
\centering
\includegraphics[width=0.48\textwidth]{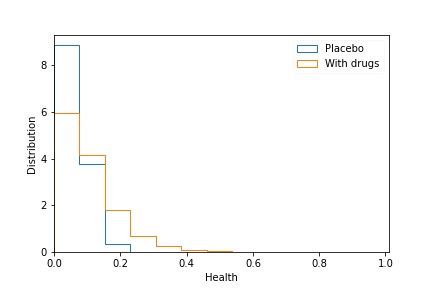}
\includegraphics[width=0.48\textwidth]{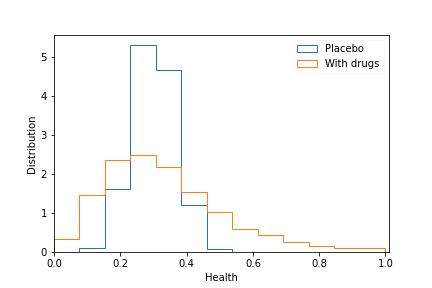}
	\caption{Distributions for functions $\H_1$ and $\H_2$, in the case of patients only and patients plus drugs, for uniformly distributed $\vec x$ and $\vec y$.}
\label{placebo-drugs}
\end{figure}

In order to normalize the following procedure and to make it more efficient, it is more appropriate to have all components in $\vec x$ and $\vec y$ to range between 0 and 1.  Therefore, any patient feature, or drug dosage should be normalized to a number in $[0,1]$.  In particular, any known to be harmful or impossible to practice drug dosage should not be taken into account, and only consider those which are scientifically and ethically possible and fit them in this range.  It is also possible to take profit of known chemical and biological facts and eventually include some suggested possible combinations into only one $y_i$ component.  This is what is usually known as designing the NN according to the requirements of the problem and we comment briefly about it in Section \ref{discussion}.

Once we have stated the above considerations, we can proceed to compare techniques RDT, NN@RDT and NNDT as follows.   Although we do not aim to propose a function $\H$, we can investigate the comparison in different and varied reasonable samples of $\H$ and extract conclusions which we can expect in data.

Along the following paragraphs we consider a scenario where patients have 5 features and there are 10 drugs to be tested in any combination.  This means that in $x_i$, $i$ goes from 1 to 5, and in $y_j$, $j$ goes from 1 to 10.  Any extension or other scenario can be easily tested, or modified, since we have open the source code of all the calculations in this work in the {\tt GIT} repo in Ref.~\cite{git}.

Along this section we present the results for two given $\H$'s, however we have also tested in different other kind of $\H$'s, as described in Appendix \ref{functions} and in \cite{git}.  We need $\H$ to satisfy a few reasonable requirements.  It should be non-linear, and it should contain features such as drug interference.  The distribution of $\H(\vec x, \vec 0)$ for random patients $\vec x$ corresponds to the outcome of patients with no drug treatment, and therefore should be mainly below some given bound, since we are assuming that an important fraction of patients do not heal without treatment.  The distribution of $\H(\vec x, \vec y)$ over $\vec x$ and $\vec y$ should yield values in the proximity of 1 (excellent health), since we are assuming that in principle exists a combination of drugs that can heal patients.  But it should also yield values in the proximity of 0 (dead), since it is reasonable to assume that there are also harmful drug combinations.  Finally, $\H$ should depend little or nothing in some of the components $x_i$ and $y_j$, since it is expected that some of the proposed drugs do not have a significant effect in the outcome of the treatment.

Using these requirements we have constructed  and tested many hypothetical $\H$.  In this work we show the results for two representative functions $\H_1(\vec x, \vec y)$ and $\H_2(\vec x, \vec y)$, whose explicit forms can be found in Appendix \ref{functions}.  Others or new functions can be tested using the tools available in Ref.~\cite{git}.   In Fig.~\ref{placebo-drugs} we have plotted their distribution for placebo treatments ($\vec y = \vec 0$) and for drug treatment by using $x_i$ and $y_j$ randomly uniformly distributed in $[0,1]$.  We stress that since only feasible drugs doses have been codified in the $y_i \in [0,1]$ variables, then this random actually means random over feasible drugs doses. An explicit and reasonable form for $\H$ (or a sample of many explicit forms) is the starting point for our analysis.

\begin{figure}[h!]
\centering
\textbf{$N=500$ patients \hskip 1.7cm $\H_1$ \hskip 1.7cm $N=2000$ patients}\par\medskip
\subfloat[]{\includegraphics[width=0.38\textwidth]{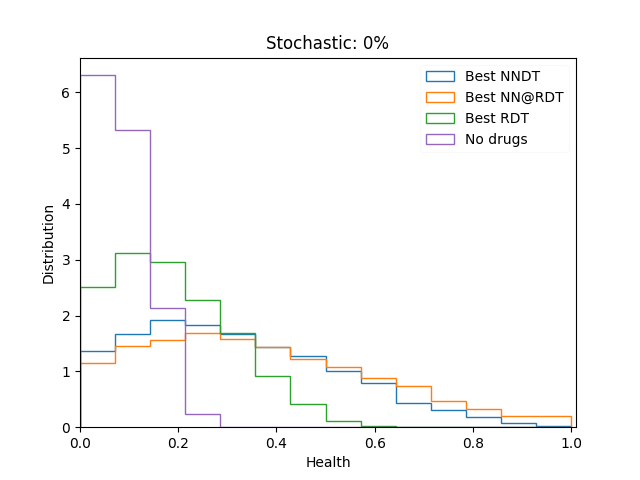}}\hspace{3mm}
\subfloat[]{\includegraphics[width=0.38\textwidth]{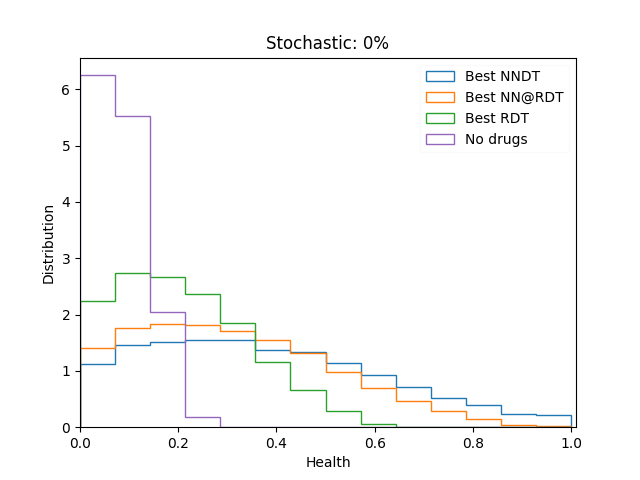}} \\
\subfloat[]{\includegraphics[width=0.38\textwidth]{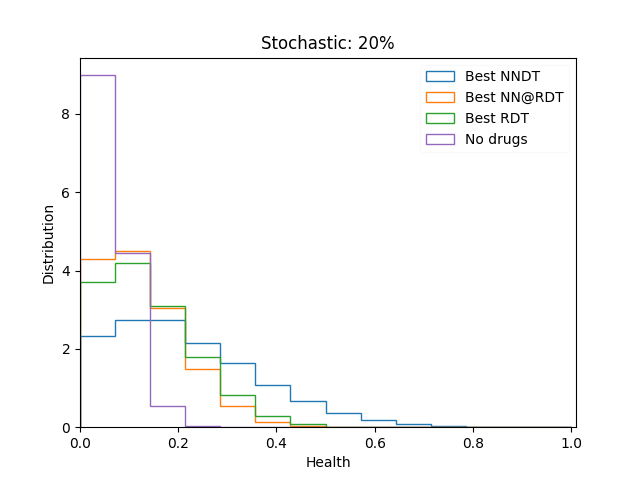}}\hspace{3mm}
\subfloat[]{\includegraphics[width=0.38\textwidth]{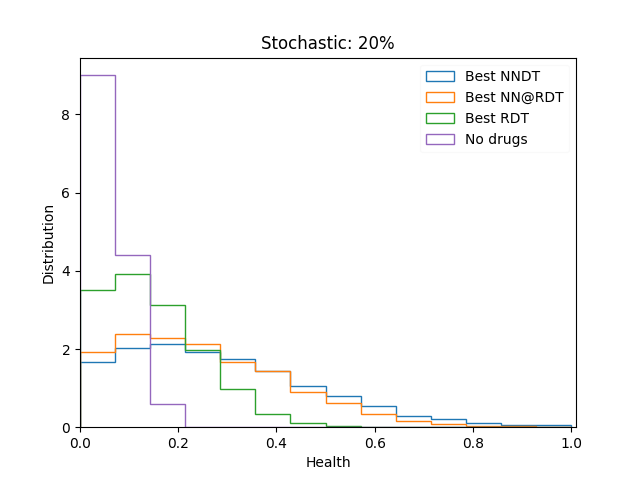}} \\
\subfloat[]{\includegraphics[width=0.38\textwidth]{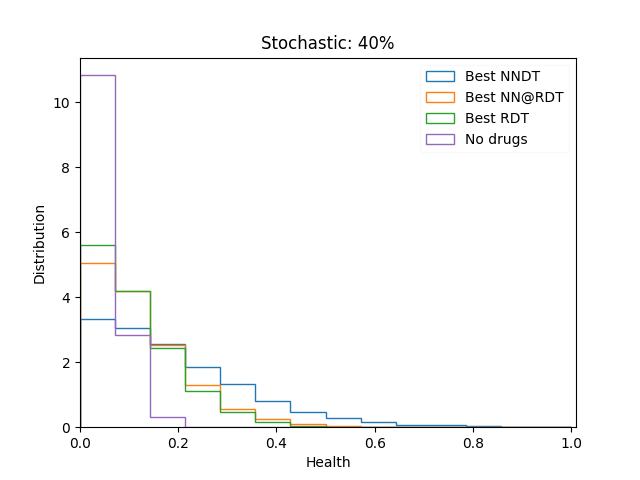}}\hspace{3mm}
\subfloat[]{\includegraphics[width=0.38\textwidth]{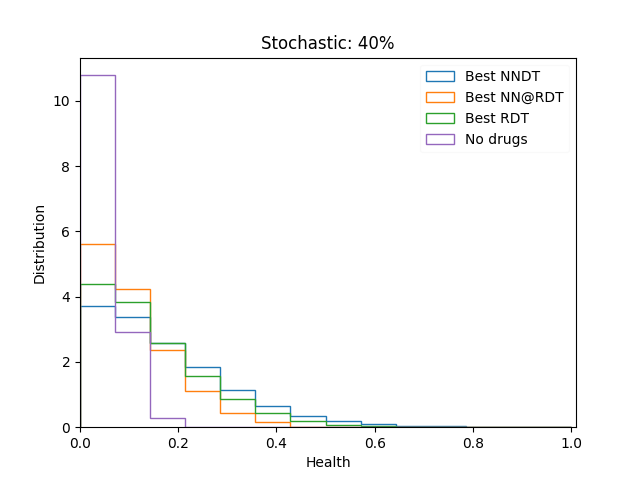}} \\
\caption{Distributions for the $\H_1$ function for different patients, with drug features fixed over each of the cases ``no-drugs'', RDT, NN@RDT and NNDT. The procedure of drug therapy discovery was done for samples of $N=500$ and $N=2000$ patients, and for values of the stochastic noise of 0\%, 20\% and 40\%. In all cases the NNDT technique (blue) performs better than the RDT (green).}
\label{moneyplots_0}
\end{figure}

\begin{figure}[h!]
\centering
%\textbf{Function $\H_2$}\par\medskip
\textbf{$N=500$ patients \hskip 1.7cm $\H_2$ \hskip 1.7cm $N=2000$ patients}\par\medskip
\subfloat[]{\includegraphics[width=0.38\textwidth]{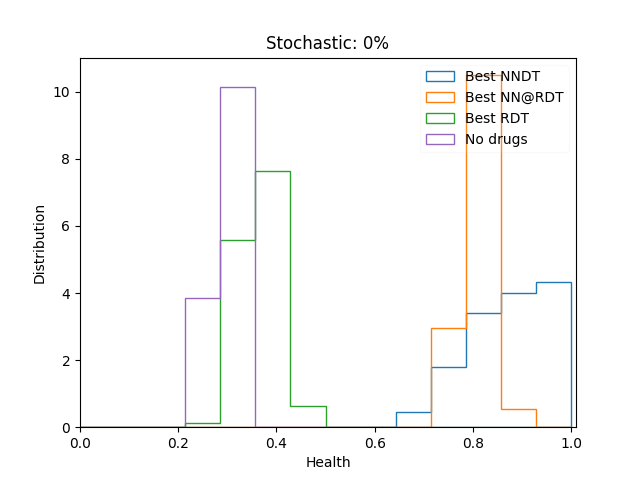}}\hspace{3mm}
\subfloat[]{\includegraphics[width=0.38\textwidth]{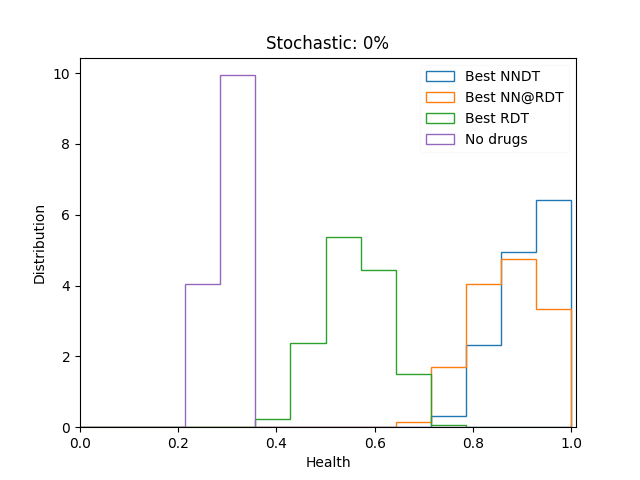}} \\
\subfloat[]{\includegraphics[width=0.38\textwidth]{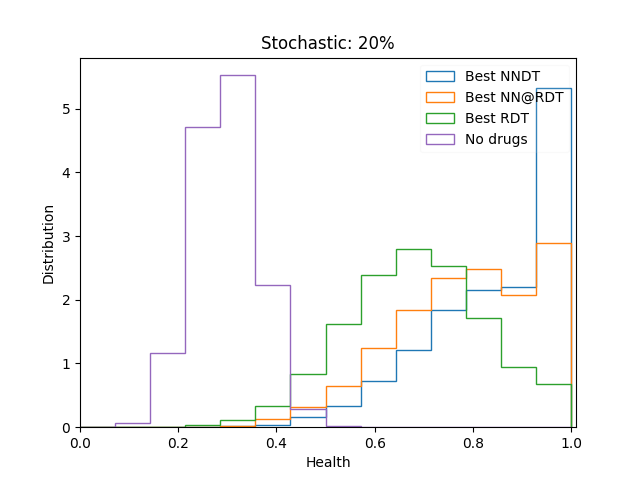}}\hspace{3mm}
\subfloat[]{\includegraphics[width=0.38\textwidth]{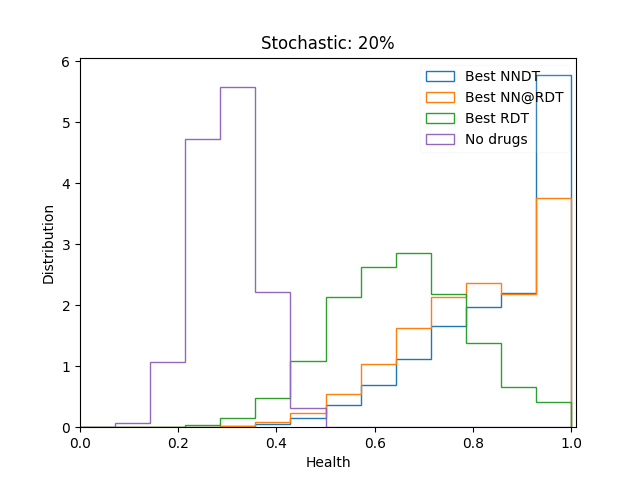}} \\
\subfloat[]{\includegraphics[width=0.38\textwidth]{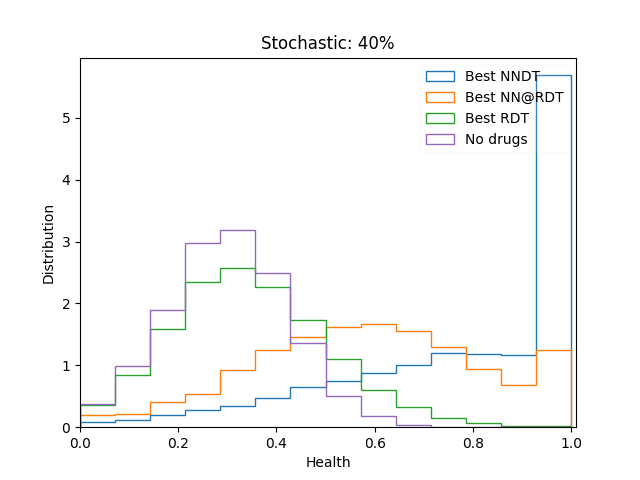}}\hspace{3mm}
\subfloat[]{\includegraphics[width=0.38\textwidth]{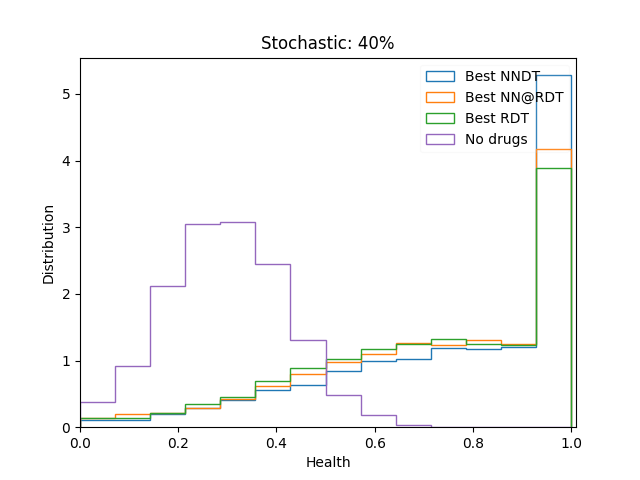}} \\
\caption{Distributions for the $\H_2$ function for different patients, with drug features fixed over each of the cases ``no-drugs'', RDT, NN@RDT and NNDT. The procedure of drug therapy discovery was done for samples of $N=500$ and $N=2000$ patients, and for values of the stochastic noise of 0\%, 20\% and 40\%.  In all cases the NNDT technique (blue) performs better than the RDT (green).}
\label{moneyplots_3}
\end{figure}

For each one of the $\H_{1.2}$ functions we explicitly perform the techniques described above RDT, NN@RDT and NNDT and compare their outcomes. This is done by computing the distribution of $\H(\vec{x},\vec{y}_{A,B,C})$ over $\vec x$, where $\vec y_{A,B,C}$  corresponds to the optimal drug found by each technique RDT, NN@RDT and NNDT, respectively.  We can interpret the mean of each one of these distributions as the average outcome of the treatment with drug $\vec{y}_{A,B,C}$, respectively. Since the function $\H$ is monotonically related to the probability of healing, if a treatment has a larger average than the others, it can be deemed the treatment with a better probability of success when averaging over types of patients. Even more, one can look at the distribution and seek drug combinations which have its shapes tilted towards $1$, avoiding bad outcomes most of the times.

We show the results of our analysis for $N=500$ and $N=2000$ patients in Figs.~\ref{moneyplots_0} and \ref{moneyplots_3} for functions $\H_1$ and $\H_2$, respectively.  In analyzing these results one should bear in mind that these functions are quite different not only in their shape, but also in their structure, as detailed in Appendix \ref{functions}.  For the Regular Drug Therapy trials, these patients are divided into groups of 100 patients each, from which half are treated with placebo. We do not consider less than 500 patients because a smaller number means too few samples of the drug space for the RDT, although we did study NN with 200 training points and found encouraging results. For each choice of $N$ and $\H$, we consider three different random noise values, which we call stochastic. Further details about the stochastic implementation can be found in Appendix~\ref{functions}. The results presented are those that are representative of the best behavior of the techniques introduced. However, the procedure is fairly stable assuming certain conditions.  Further details about the NN hyperparameters choice and its performance can be found in Appendix~\ref{NN}.

From both Figures we see that the NN-reliant techniques perform better than RDT, finding both a higher average output and a more probable favorable output than RDT. Most of the times, NNDT performs better than NN@RDT. However, NN@RDT which performs better than RDT, in some cases --not shown-- reaches the NNDT level or slightly better.  While both cases are favorable, we also see that $\H_1$ and $\H_2$ model different kind of functions: $\H_1$ yields a more evenly distributed outcome for every technique, while $\H_2$ yields a `lumped' result for each technique. In both cases, the NN is found to provide a good reconstruction of the underlying function $\H$, yielding Spearman rank correlation coefficients ranging $R \sim 0.7 - 0.95$ depending on $N$ and the stochastic noise parameter, as detailed in Appendix \ref{NN}

\section{Discussion}
\label{discussion}

The work and results presented so far have different possible improvements and followups.  Many of them can be taken on directly from the available open source code in {\tt GIT} repo \cite{git}, where all the tools to reproduce and expand the article results are publicly available.

Along the investigation we have found that the NN architecture is important to achieve good results.  The NN best architecture is dependent on the number of cases analyzed ($N$) since, as $N$ is reduced, the architecture should be reduced as well to avoid overfitting.  Also, we find that the higher is the complexity and non-linearity of $\H$, the deeper (more layers) should be the NN.   Further investigations and trials in understanding the best NN architecture design would play in favor of more solid results considering a real scenario.

This work could be complemented with another ML technique to best choose the variables $\vec x$ and $\vec y$.  In particular one could run an auto-encoder on the whole set of these variables and let the algorithm reduce the variables into more relevant ones.  These latter could be used as input in the NN used along this work and eventually optimize times and performance of the network.

An important point that can be extracted from this work is related to the function $\H$ and its true value.  Although along the article we have tested general samples of $\H$ for the sake of finding general behaviors and properties, having true expected features coming from the biology and physician expertise could provide more realistic forecasts and more focused NNs.   Moreover, on the way around, one could attempt to reconstruct how $\H$ depends on its variables through the NN and compare it to molecular and biological models to learn from its description.

The NNDT presented seems to have, at least from the mathematical point of view, better prospects than RDT.  However, this NNDT could still be improved as follows.  Once the NN has been trained with the real patients data, upon the need to apply the best drug therapy to a new patient, one could use the NN to design the best drug therapy {\it customized} for this new patient.  The procedure is to set the $\vec x$ fixed to the new patient features, and vary at random the drugs $\vec y$ until finding the maximum outcome of the NN.  Such a drug combination $\vec y_C$ would be a customized drug therapy for the patient features.  We have verified that this technique is as good or better than NNDT, however to obtain the corresponding distributions of this customized NN Drug Therapy within the presented framework requires large CPU resources.  We plan to further study this possibility and eventually include them in a next update.

At last, we mention similarities and differences between our work and the one in Ref.~\cite{kaptein}. The main similarity in both studies is the mathematical and phenomenological setup of the problem in contrast to the usual approach.  On the other hand, Ref.~\cite{kaptein} proposes a Bayesian additive regression tree model based on sequential experimentation, whereas our approach is based on a NN that analyzes the whole patients dataset at once, since time is crucial for our objectives.  Our scheme proposes to work with an explicit function $\H$ which allows us to study in more detail the differences between RDT and NN approaches within this framework.  We consider that both works complement each other and push forward the same idea of considering replacing RDT for a more sophisticated Machine Learning algorithm.

\section{Conclusion}
\label{conclusions}

We have investigated, from a mathematical point of view, how new available Machine Learning techniques could improve the efficacy and reduce times of clinical trials in finding the best drug therapy upon a new unknown disease.  

The main point of our analysis consists in replacing usual placebo-controlled clinical trial techniques of a fraction of patients treated by a given specific drug and the other patients by placebo, by all patients treated with a variety of different possible drug therapies each one.  We have shown that a Neural Network can assimilate the variety of data and expected fluctuations without need of large number of patients under the same treatment.  

To compare the prediction of a trained Neural Network against usual clinical trials techniques we have implemented a phenomenological approach in which we model the evolution of a patient with specific features from beginning to end of treatment as function of the drug combination received.  This modeling does not rely on any physiological, biological, nor molecular behavior or interaction, but instead just on the outcome of the patient as a number in $[0,1]$ related to the patient health at the end of treatment.  

Our findings show that a Neural Network Drug Therapy (blue line in Figs.~\ref{moneyplots_0} and \ref{moneyplots_3}) performs always better than a Regular Drug Therapy (green line).  The strength of the argument resides in that this result holds regardless of the specific $\H$ used.

We also discuss along the article potential developments and improvements that could be done from this result.  Among them, we propose that using such a Neural Network technique could be used to test any modeling as described above, or that the Machine Learning technique could provide a still better customized drug therapy to each specific patient.  Further work is needed in many of these fronts.  We also observe that the results presented in this work are useful as well for other disciplines in which experimental techniques require a test and control group in different scenarios to understand different behaviors and/or patterns.

We understand this work as a proof-of-concept of the presented idea.  Further investigations along the biopharma and physician sides would be required to explore whether some of the ideas here proposed could be taken to a real scenario.  In such a case, this could provide an important step in accelerating drug therapy discovery in important health issues, such as for instance the COVID-19 pandemic.

\newpage
\appendix

\section{Functions $\H$}
\label{functions}

In this work we modeled the behavior of a practically intractable physical system with a function $\H(\vec{x},\vec{y})$ which has a fairly general set of hypothesis described in Sections~\ref{problem},\ref{pheno}. To perform a practical test of the techniques proposed we use a set of functions $\H(\vec{x},\vec{y})$ which aim to capture the following hypotheses:

\begin{itemize}
	\item It is a multi-variable (highly) non-linear function.
	\item It can take values in $[0,1]$.
	\item It has a stochastic component.
	\item $\H(\vec{x},0)$ should be tilted towards 0 to represent the no-drugs expected outcome.
        \item $\H(\vec{x},\vec{y})$ may reach higher or lower values than $\H(\vec{x},0)$, representing those drugs $\vec y$ that have a positive or harmful effect for health, respectively. 
          \item It can contain features such as drug interference or cancellation. For example having a dependence like $x_k(y_i - y_j)$ allows for such a behavior.
\end{itemize}
Results presented in this work were done using two different forms for the $\H$ function, $\H_1$ and $\H_2$. Their differences lie in the way they are constructed, which is discussed in the following paragraphs.

\subsection*{Function $\H_1$} 
$\H_1$ is factorized into a patient-specific part $P$, a drug-specific part $d_0$, a function of both patient and drugs $d_1$, and the stochastic piece $S$ depending on a parameter $\eta$.
\be
\H_1(\vec{x},\vec{y}) = P(\vec{x})\, d_0(\vec{y}) \, d_1(\vec{x},\vec{y})\, S_\eta  
\te 
with

\al
P(\vec{x}) =& \Bigg\vert \frac{\sum{(\alpha_k - 2 \alpha_k x_k)}}{\sum{\alpha_k}}\Bigg\vert^{\psi} \nn \\
d_0(\vec{y}) =& {\rm exp}(-|\dd_1 y_{9} + \dd_2 y_1 + \dd_3 y_2 + \dd_4 y_3 + \dd_5 y_4| \nn \\
& + |\dd_6 y_{5} + \dd_7 y_2 + \dd_8 y_{6} + \dd_9 y_{7} | \nn \\
&  + |\dd_{10} y_{8} + \dd_{11} y_{9} + \dd_{12} y_1 + \dd_{13} y_{10}|/10)  \nn \\
d_1(\vec{x},\vec{y})=& 1 + 0.1 \sin(\ddd_1 y_1 - |(\ddd_2 y_4 + \ddd_3 y_{5} + \ddd_4 y_{6})(\ddd_5 x_1 y_3 - \ddd_6 x_3 y_{9})|) \nn\\ 
\fal
The patient function has a dependence on a coefficient $\psi$ that controls how easily patients can heal, in the absence of drugs. For $\psi<1$ this function leans towards higher values, and for $\psi>1$, towards zero.
The drug function $d_0$ has an exponential dependence over a combination of drug parameters $y_i$. The part $d_1$ adds an oscillation that depends on a specific combination of drugs and patient features, allowing here for certain combinations to conspire into an increase or a reduction of the whole value of the function. For example in the second term inside the $\sin$ function we see that there is a factor that contains the combination $x_1 y_3 - x_3 y_9$, for which takes into account drug interference between $y_3$ and $y_9$, weighted by the features $x_1$ and $x_3$.
The rest of the functions parameters $\alpha_i, \dd_i, \ddd_i$ are selected at random, ranging in $\alpha_i \in [0,10]$, $\dd_i,\ddd_i \in [-1,1]$. For each specific set of values of the parameters, we have a certain function $\H_1$. Thus, the above formulas describe a family of functions $\H_1$. We checked that for several values of the parameters the distributions of $\H_1$ with and without drugs are sufficiently separated, as in Fig.~\ref{pheno}. We then fixed the parameters' values to reproduce such behavior.
Regarding the noise factor $S$, we use a Gaussian function centered at 1, with a standard deviation $\eta$ and independent of $\vec x$ and  $\vec y $. As we want the output of the function to be in the range [0,1], we have to do two things. First, as it cannot take negative values, we take S to be the absolute value of the Gaussian centered around 1. Then, we calculate the maximum value attained by the part $P\cdot d_0 \cdot d_1$, and take into account possible fluctuations up to two standard deviations $\eta$. We then divide the function by this value  $\H_1^{(0)} \equiv \max_{\vec{x},\vec{y}}( P(\vec{x}) d_0(\vec{y}) d_1(\vec{x},\vec{y})) (1+2 \eta)$. Then we bound the value of the function to be lower or equal than one. That is, if a certain fluctuation of the noise is outside of two standard deviations and the function evaluates to something higher than 1, we take it to be equal to 1.

\subsection*{Function $\H_2$}
In the case of function $\H_2$, it is not factorizable, but has a more compact form
\be
\H_2 = \frac{1}{15} \Bigg\vert x_1+(y_1 + 3 y_2 - y_3)(x_5-x_3)+ \sinh(y_7-y_6)- 5\, e^{-(y_9 -y_3)} \Bigg\vert S_\eta 
\te 
Once again, it has the features described above, the nonlinearities, together with drug interference/cancellation. The distributions of this function with and without placebo can be seen in Fig.~\ref{pheno}. What can be seen is that in this case the distribution with drugs takes smaller values than the distribution without them, that is, there are combinations of drugs that are harmful for the patients.

\section{Neural Network Performance and Validation}
\label{NN}

The results presented in Fig.~\ref{moneyplots_0} and \ref{moneyplots_3} correspond to a particular choice of hyperparameters for the Neural Networks-reliant techniques.   The validity of the NN@RDT and NNDT techniques, needs to take into account not only how good is $\H(\vec x, \vec y_{B,C})$, but how well does the NN perform at reconstructing this truth-level distribution. 

For the task of building and training the Neural Networks we use Python programming language and resort to the {\tt Keras} package \cite{Keras}, for each of the functions we scanned over different architectures and other hyperparameters. 
The Neural Networks considered were composed of fully connected layers, with RELU activation in all of the units, with a single neuron in the final layer. We turn off the bias in most of the hidden layers to prevent a shift in the output of the NN.  As the goal at hand is a regression problem --we want to interpolate the $\H$ function given a certain number of sample points--, we used a mean squared error loss function. Each of the $\H$ functions is different in complexity and, as such, it would be recommendable to correspondingly choose the number of epochs, the learning rate and the number of neurons.  As we need to fit nonlinear functions, we consider deep networks with a number of hidden layers between 4 and 6. For the results in this article, the used architecture consists of 8 hidden layers with neuron numbers of $(100,32,10,10,10,10,10,10)$ respectively.  As we need to avoid overfitting, we have to select architectures with a number of parameters lower than the number of degrees of freedom present in the input data. For example, for $N=500$, 5 patient features and 10 drug features, we have $7500$ degrees of freedom, for which the complexity of the network cannot be too large. The overfitting of the NN can be checked by plotting the loss function over the training and validation sets.  In any case, we have worked in all NN with a dropout of 10\% to reduce overfitting.

\begin{figure}[h]
\centering
\textbf{Case 1: 500 and 2000 patients}\par\medskip
\subfloat[]{\includegraphics[width=0.48\textwidth]{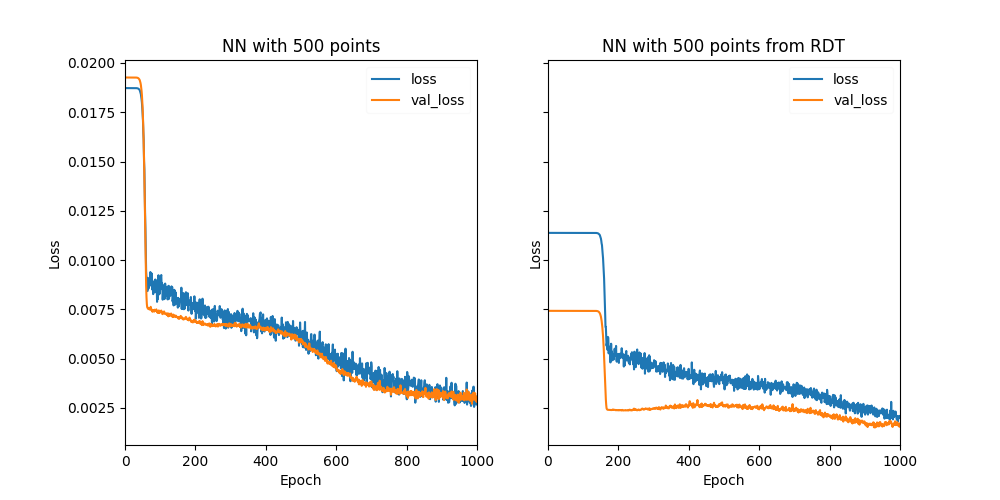}}\hspace{3mm}
\subfloat[]{\includegraphics[width=0.48\textwidth]{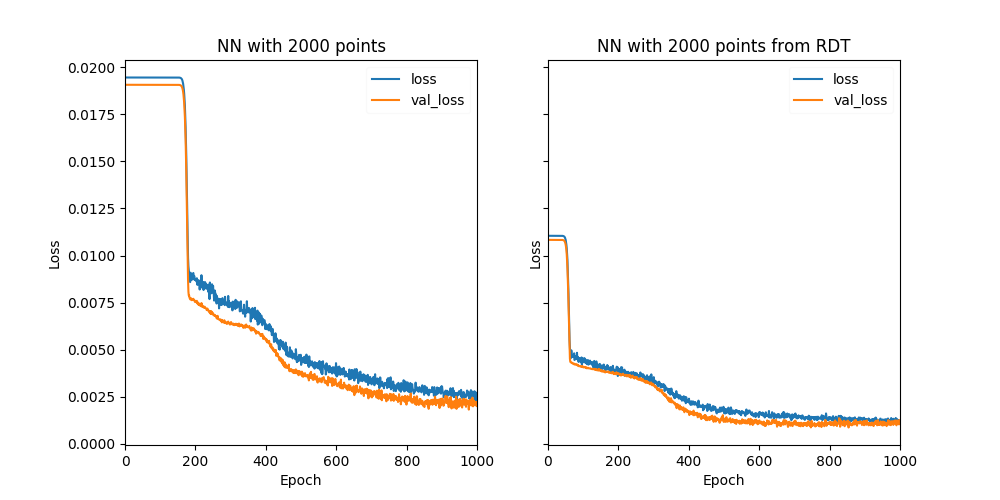}} \\
\subfloat[]{\includegraphics[width=0.48\textwidth]{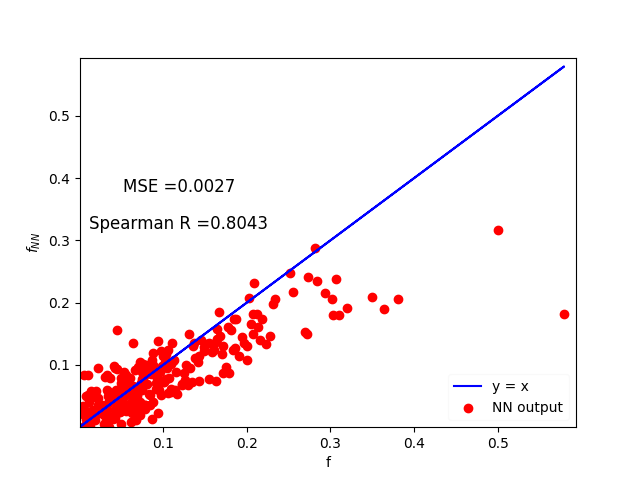}}\hspace{3mm}
\subfloat[]{\includegraphics[width=0.48\textwidth]{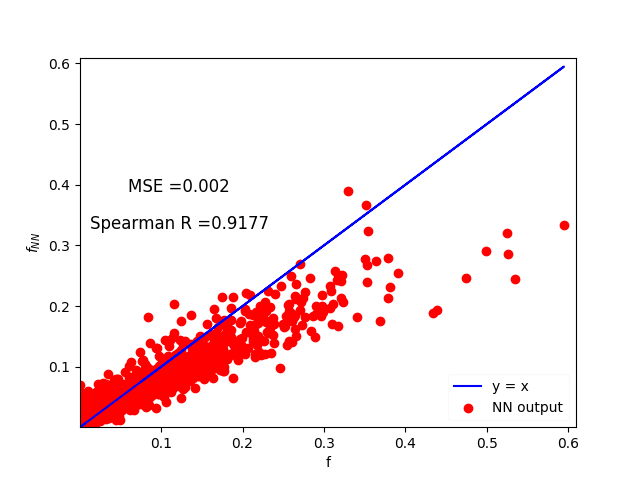}} \\
\subfloat[]{\includegraphics[width=0.48\textwidth]{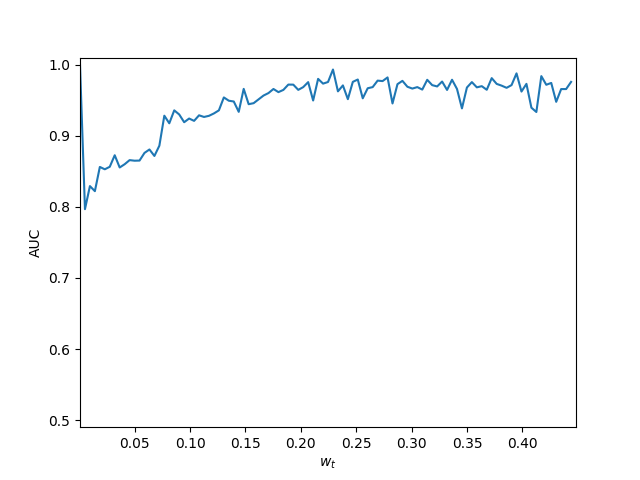}}\hspace{3mm}
\subfloat[]{\includegraphics[width=0.48\textwidth]{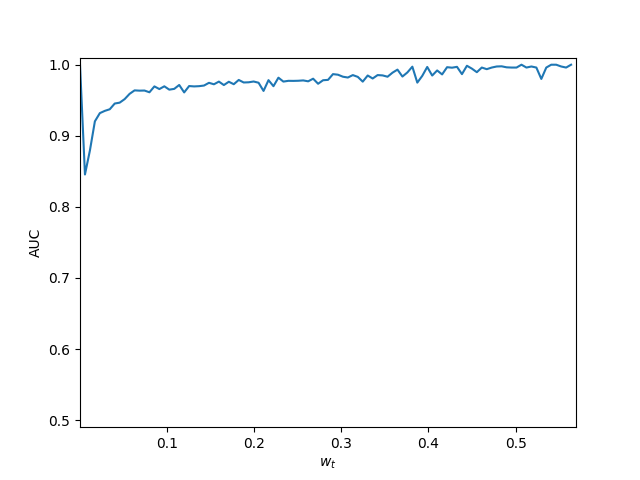}} \\
	\caption{Different measures of the goodness of the NN fit, for function $\H_1$, for both $N=500$ and $N=2000$ train sizes, with a stochastic noise of $\eta$ = 0.2.}
\label{NNmoneyplots_0}
\end{figure}

\begin{figure}[h]
\centering
\textbf{Case 2: 500 and 2000 patients}\par\medskip
\subfloat[]{\includegraphics[width=0.48\textwidth]{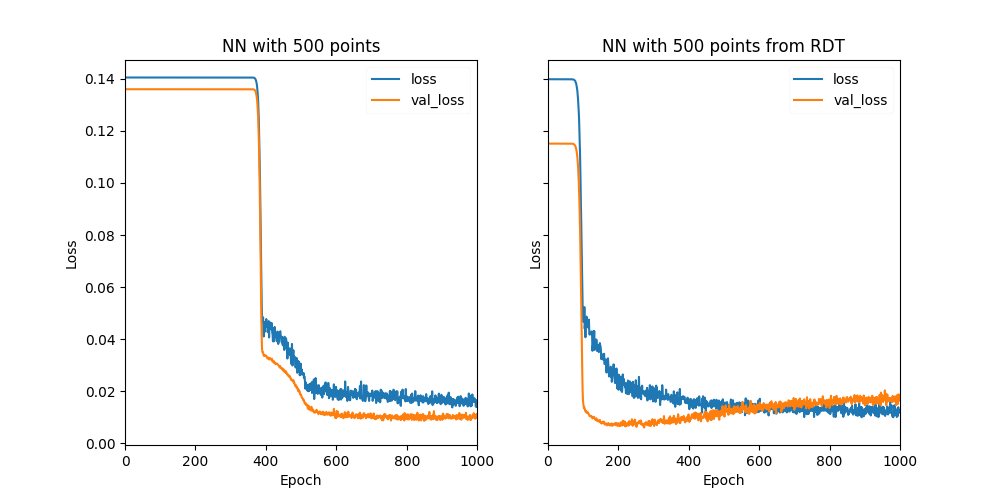}}\hspace{3mm}
\subfloat[]{\includegraphics[width=0.48\textwidth]{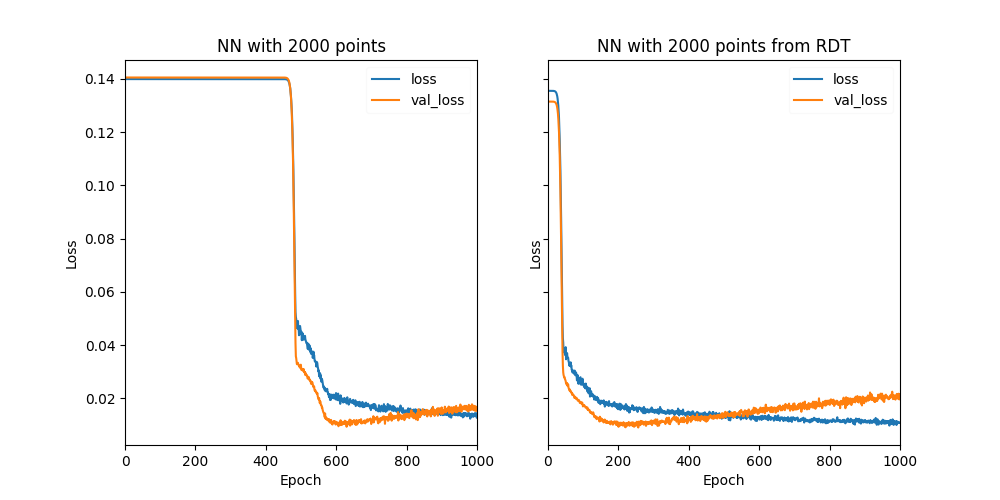}} \\
\subfloat[]{\includegraphics[width=0.48\textwidth]{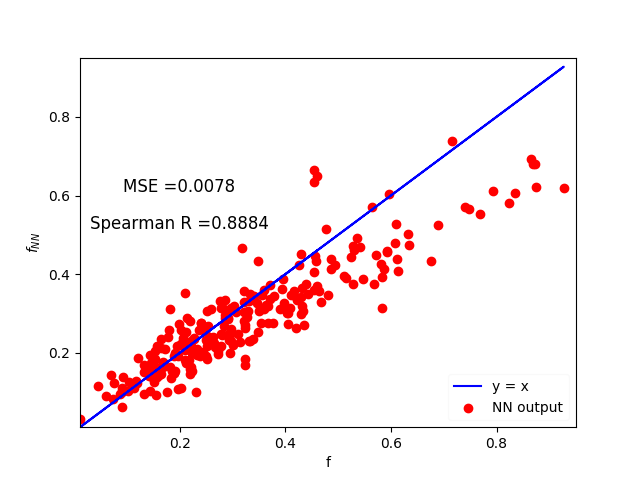}}\hspace{3mm}
\subfloat[]{\includegraphics[width=0.48\textwidth]{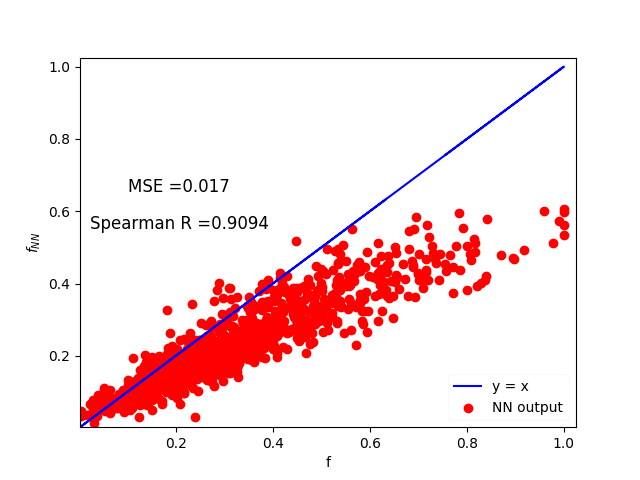}} \\
\subfloat[]{\includegraphics[width=0.48\textwidth]{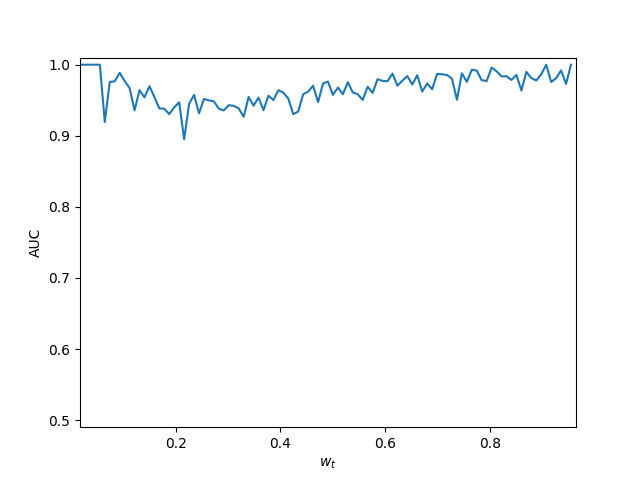}}\hspace{3mm}
\subfloat[]{\includegraphics[width=0.48\textwidth]{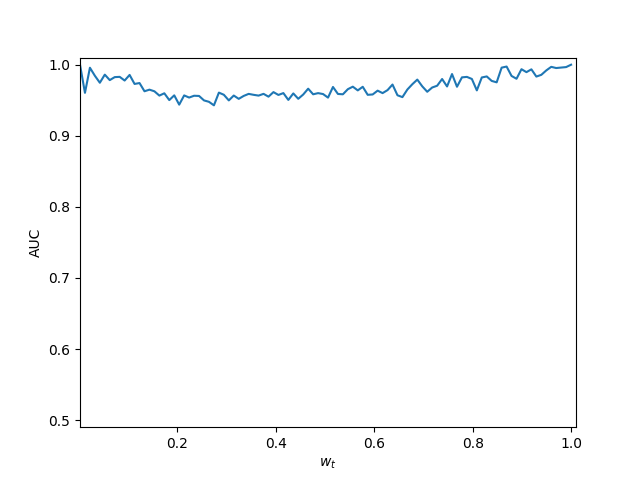}} \\
	\caption{Different measures of the goodness of the NN fit, for function $\H_2$, for both $N=500$ and $N=2000$ train sizes, with a stochastic noise of $\eta$ = 0.2. }
\label{NNmoneyplots_3}
\end{figure}

We show the performance of each NN architecture chosen for a given $\H$ and $N$ in three plots in Fig.~\ref{NNmoneyplots_0},~\ref{NNmoneyplots_3}, the first of which is the Loss function. The remaining two plots use a test set of previously unseen points to compare the predicted values of $\H$ against the true values, which we can denote as ($\H(x,y),\H_{NN}(x,y)$). We plot this set of points, comparing it to the $\H_{NN}=\H$ curve, and we calculate the mean squared error and the Spearman's rank correlation coefficient. These two are used as a measure of how adequate the hyperparameters are for each $\H$ and number of patients $N$ to reproduce the function $\H$.

Another test to see how well does $\H_{NN}$ reproduce the behavior of $\H$ over the test data is to use $\H$ to select whether a patient heals or not and test if $\H_{NN}$ categorizes the data in the same way. This can be done by setting a threshold in $\H$ (which we call $w_t$) to label the data and then computing the Area Under Curve (AUC) when trying to reproduce these labels with $\H_{NN}$. While computing the mean squared error treats the fidelity point by point, this method checks whether $\H_{NN}$ reproduces the true function behavior over the whole data by yielding a consistently high AUC for each $w_{t}$. A lower AUC for a region of $w_{t}$ means that the NN is not able to capture the true variation of outcomes over some data region.

We see that both cases yield a fairly good Spearman correlation coefficient while also giving a quite loyal AUC behavior, even when considering a stochastic component. If we increase the noise, the performance worsens but not to an intolerable level. From these results we assert that we can trust the above NN framework to describe the Health distributions.

\bibliographystyle{JHEP}
\bibliography{biblio}
\end{document}